\documentclass[]{aastex631}

\newcommand{\T}{$T_{\rm ref}$}

\newcommand\redsout{\bgroup\markoverwith{\textcolor{red}{\rule[0.5ex]{2pt}{0.4pt}}}\ULon}

\usepackage{amsmath}
\usepackage{graphicx}
\usepackage{ulem}
\usepackage{natbib}
\usepackage{subdepth}

\begin{document}

\title{Relation between the keV-MeV and TeV emission of GRB 221009A and its implications}

\correspondingauthor{Shao-Lin Xiong, Haoxiang Lin, Ming-Yu Ge, Zhuo Li}
\email{xiongsl@ihep.ac.cn, haoxiang@pku.edu.cn, gemy@ihep.ac.cn, zhuo.li@pku.edu.cn}

\author{Yan-Qiu Zhang}
\affiliation{Key Laboratory of Particle Astrophysics, Institute of High Energy Physics, Chinese Academy of Sciences, Beijing 100049, China}
\affiliation{University of Chinese Academy of Sciences, Chinese Academy of Sciences, Beijing 100049, China}

\author{Haoxiang Lin*}
\affiliation{Kavli Institute for Astronomy and Astrophysics, Peking University, Beijing 100871, China}

\author{Shao-Lin Xiong*}
\affiliation{Key Laboratory of Particle Astrophysics, Institute of High Energy Physics, Chinese Academy of Sciences, Beijing 100049, China}

\author{Zhuo Li*}
\affiliation{Kavli Institute for Astronomy and Astrophysics, Peking University, Beijing 100871, China}
\affiliation{Department of Astronomy, School of Physics, Peking University, Beijing 100871, China}

\author{Ming-Yu Ge*}
\affiliation{Key Laboratory of Particle Astrophysics, Institute of High Energy Physics, Chinese Academy of Sciences, Beijing 100049, China}

\author{Chen-Wei Wang}
\affiliation{Key Laboratory of Particle Astrophysics, Institute of High Energy Physics, Chinese Academy of Sciences, Beijing 100049, China}
\affiliation{University of Chinese Academy of Sciences, Chinese Academy of Sciences, Beijing 100049, China}

\author{Shu-Xu Yi}
\affiliation{Key Laboratory of Particle Astrophysics, Institute of High Energy Physics, Chinese Academy of Sciences, Beijing 100049, China}

\author{Zhen Zhang}
\affiliation{Key Laboratory of Particle Astrophysics, Institute of High Energy Physics, Chinese Academy of Sciences, Beijing 100049, China}

\author{Shuang-Nan Zhang}
\affiliation{Key Laboratory of Particle Astrophysics, Institute of High Energy Physics, Chinese Academy of Sciences, Beijing 100049, China}

\author{Li-Ming Song}
\affiliation{Key Laboratory of Particle Astrophysics, Institute of High Energy Physics, Chinese Academy of Sciences, Beijing 100049, China}

\author{Chao Zheng}
\affiliation{Key Laboratory of Particle Astrophysics, Institute of High Energy Physics, Chinese Academy of Sciences, Beijing 100049, China}
\affiliation{University of Chinese Academy of Sciences, Chinese Academy of Sciences, Beijing 100049, China}

\author{Wang-Chen Xue}
\affiliation{Key Laboratory of Particle Astrophysics, Institute of High Energy Physics, Chinese Academy of Sciences, Beijing 100049, China}
\affiliation{University of Chinese Academy of Sciences, Chinese Academy of Sciences, Beijing 100049, China}

\author{Jia-Cong Liu}
\affiliation{Key Laboratory of Particle Astrophysics, Institute of High Energy Physics, Chinese Academy of Sciences, Beijing 100049, China}
\affiliation{University of Chinese Academy of Sciences, Chinese Academy of Sciences, Beijing 100049, China}

\author{Wen-Jun Tan}
\affiliation{Key Laboratory of Particle Astrophysics, Institute of High Energy Physics, Chinese Academy of Sciences, Beijing 100049, China}
\affiliation{University of Chinese Academy of Sciences, Chinese Academy of Sciences, Beijing 100049, China}

\author{Yue Wang}
\affiliation{Key Laboratory of Particle Astrophysics, Institute of High Energy Physics, Chinese Academy of Sciences, Beijing 100049, China}
\affiliation{University of Chinese Academy of Sciences, Chinese Academy of Sciences, Beijing 100049, China}

\author{Wen-Long Zhang}
\affiliation{Key Laboratory of Particle Astrophysics, Institute of High Energy Physics, Chinese Academy of Sciences, Beijing 100049, China}
\affiliation{School of Physics and Physical Engineering, Qufu Normal University, Qufu, Shandong 273165, China}

\begin{abstract}
Gamma-ray bursts (GRBs) are believed to launch relativistic jets, which generate prompt emission by internal processes, and produce long-lasting afterglows by driving external shocks into surrounding medium. However, how the jet powers the external shock is poorly known. The unprecedented observations of the keV-MeV emission with GECAM and the TeV emission with LHAASO of the brightest-of-all-time GRB 221009A offer a great opportunity to study the prompt-to-afterglow transition and 
the impact of jet on the early dynamics of external shock. In this letter, we find that the cumulative light curve of keV-MeV emission could well fit the rising stage of the TeV light curve of GRB 221009A, with a time delay, $4.45^{+0.26}_{-0.26}$\,s, of TeV emission. Moreover, both the rapid increase in the initial stage and the excess from about \T+260\,s to 270\,s in the TeV light curve are tracking the light-curve bumps in the prompt keV-MeV emission. The close relation between the keV-MeV and TeV emission reveals the continuous energy-injection into the external shock. Assuming an energy-injection rate exactly following the keV-MeV flux of GRB 221009A, including the very early precursor, we build a continuous energy-injection model where the jet Lorentz factor is derived from the TeV time delay, and the TeV data is well fitted, with the TeV excesses interpreted by inverse Compton (IC) scatterings of the inner-coming prompt emission by the energetic electrons in external shock. 

\end{abstract}


\section{Introduction}

Gamma-ray bursts (GRBs) are widely recognized as the most violent explosions in the universe \citep{Zhangbing_GRB}. Based on the duration distribution of their prompt emission, they can be classified 
\citep[e.g.][]{1993ApJ...413L.101K, 2009ApJ...703.1696Z}
as short GRBs (SGRBs or Type I GRBs) produced by binary compact star mergers \citep{Short_GRB} and long GRBs (LGRBs or Type II GRBs) originated from massive star core collapses \citep{Long_GRB}. Either type of GRBs forms a central engine and emits a pair of ultra-relativistic and collimated jets. Prompt emission in the keV-MeV energy range could be generated by internal or magnetic dissipation of jet energy \citep[e.g.,][]{Rees:1994nw, zhang2011}. Afterglow arises from the external shock formed from the interaction between the jet and the surrounding medium \citep{1997ApJ...476..232M} \citep[see also][for review]{gao2013NewAR}. 

In the multi-wavelength astronomy era, the brightest GRB to date, GRB 221009A, has attracted numerous observations of many facilities spanning from radio to TeV.
Among all observations in the keV-MeV band where the GRB radiates most of its energy, GECAM-C obtained the most accurate temporal and spectral measurement of the prompt emission without data saturation effect \citep{HXMT-GECAM:GRB221009A}. LHAASO obtained the very early coverage and high-statistics light curve in the TeV band that is dominated by external forward-shock emission \citep{lhaaso2023tera}. 
Historically, there have been some GRBs with prompt keV-MeV emission and GeV/TeV observations, e.g., GRB 130427A \citep{ackermann2014fermi} and some other Fermi-LAT detected GRBs \citep{2011MNRAS.415...77M}. However, the past observations can hardly pin down the relationship between the prompt keV-MeV emission and TeV emission, because the onset of the external-shock emission was not clearly observed. Thus, the observations of GRB 221009A provide a precious opportunity to investigate this relation and the dynamics of the external shock soon after the explosion begins.

\section{Observations}
GECAM (Gravitational wave high-energy Electromagnetic Counterpart All-sky Monitor) is a dedicated gamma-ray telescope network to monitor high energy transients, such as GRBs (e.g. \cite{HXMT-GECAM:GRB221009A,GRB230307A_sunhui,mini_jet_yishuxu}), Soft Gamma-ray Repeaters (e.g. \cite{Minimum_Variation_Timescales_xiao}), Solar Flares (e.g. \cite{quasi-periodic_SF_zhaohs}), X-ray Binary bursts (e.g. \cite{XRBs}), Terrestrial Gamma-Ray Flashes and Terrestrial Electron Beams (e.g. \cite{TGF_TEB_zhaoyi}). The first two micro-satellites, GECAM-A and GECAM-B, were launched on December 10, 2020 \citep{li2022technology}.

As the third instrument of the GECAM series, GECAM-C (also known as HEBS) was launched onboard the SATech-01 satellite on July 27, 2022 \citep{ZHANG2023168586}. 
GECAM-C is equipped with 12 Gamma-ray detectors (GRDs)\citep{an2022design} and two Charge particle detectors (CPDs) \citep{xu2022design}, distributed on two domes installed on the opposite sides of the satellite. The GECAM-C detectors have two electronic readout channels for different energy detection ranges, i.e. high-gain (HG) and low-gain (LG). In addition, on-ground calibration tests and in-flight cross-calibration have been performed, which show that the GECAM-C has good performance \citep{zheng:2023ground_calibration, Zhangyq:2023cross_calibration}. 

GECAM-C accurately measured GRB 221009A without any data saturation throughout this burst \citep{GCN.32751} \footnote{\url{https://gcn.gsfc.nasa.gov/gcn3/32751.gcn3}}, thanks to the dedicated designs in the detector and electronics systems and the special working mode for the high latitude region \citep{HXMT-GECAM:GRB221009A}. With the joint observation of GECAM-C, Insight-HXMT and Fermi/GBM, unprecedented discoveries of the emission line up to 37 MeV and its power-law time evolution, intriguing properties of the early afterglow light curve and spectrum, and the remarkable behavior of the jet break have been made \citep{HXMT-GECAM:GRB221009A, Emission_Lines, Afterglow_zhengchao}.

\begin{figure}[!htb]
  \centerline{
      \includegraphics[width=0.55\columnwidth]{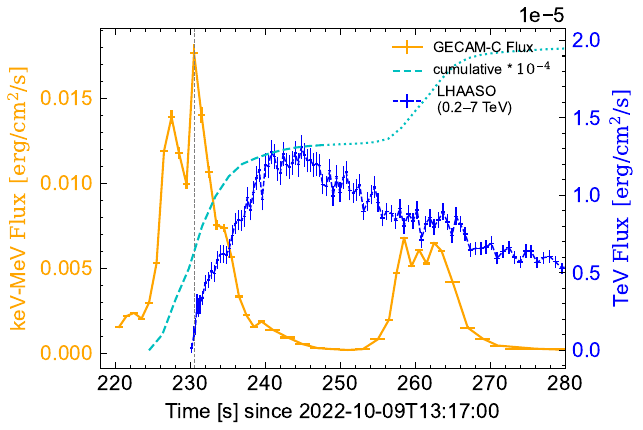}
      }
  \caption{Light curves of GRB 221009A in the keV-MeV band measured with GECAM-C (orange) \citep{HXMT-GECAM:GRB221009A} and TeV band measured by the WCDA of LHAASO (blue) \citep{lhaaso2023tera}. The orange color represents the keV-MeV flux light curve of the bolometric emission in the energy range of 1-10000\,keV/(1+z). The cyan dashed line is the cumulative curve of the keV-MeV emission (orange). The cumulative curve is scaled down by $10^{-4}$ for comparison with TeV light curve.}
  \label{fig:flux}
\end{figure}

The keV-MeV light curve measured by GECAM-C and the TeV light curve by LHAASO are shown in Figure \ref{fig:flux}, where \T=2022-10-09T13:17:00.000 (UTC) is used as the reference time for convenience. The energy range of the keV-MeV flux is 1-10000\,keV/(1+z), which is also used to calculate the isotropic bolometric emission energy. The redshift $z=0.151$ is adopted for this burst. There are two bumps in the most bright episode (dubbed main burst) of this burst from \T + 220\,s to 280\,s. The vertical grey dotted line denotes the time of the highest peak in the keV-MeV prompt emission measured with GECAM-C. The cyan dashed line represents the cumulative curve of the keV-MeV flux light curve, with the long dashed line and dotted line representing the cumulative curve of the first bump (from \T + 225\,s to 240\,s) and the second bump (from \T + 250\,s to 270\,s) of the main burst, respectively.

\section{Data analysis and results}\label{sec:data}
\begin{figure}[h]
  \centerline{
      \includegraphics[width=0.53\columnwidth]{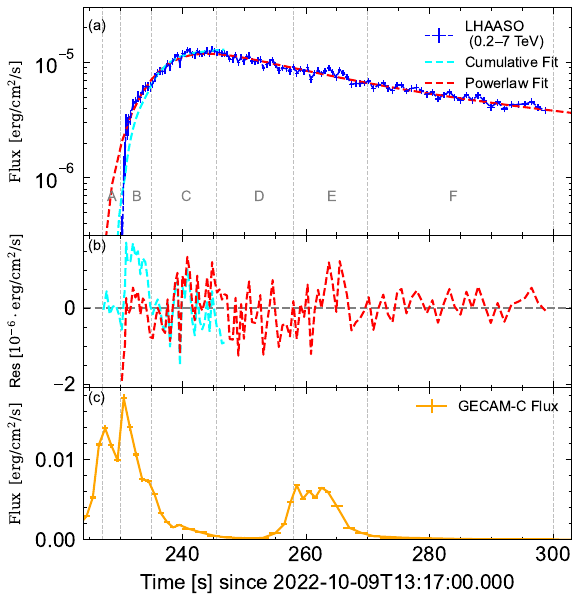}
      \includegraphics[width=0.5\columnwidth]{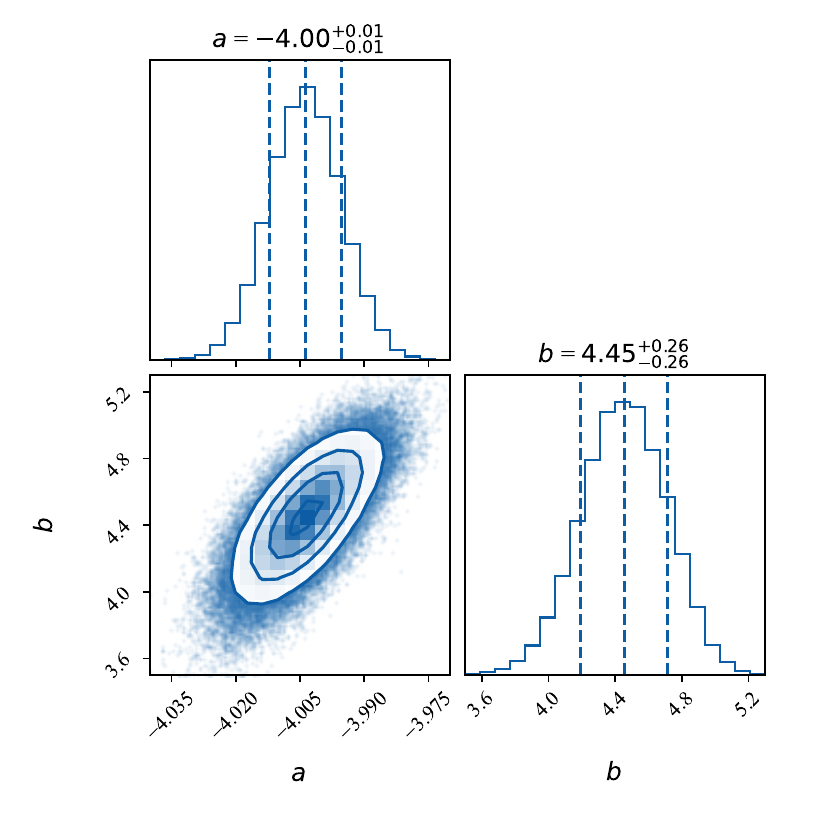}
      }
  \caption{Results of fitting the TeV light curve using the cumulative curve of the keV-MeV emission. The fitted time period is \T+(235.0,245.5)\,s (i.e. time interval C), as shown in panel (a). In panel (a), the blue data points are the TeV light curve with LHAASO and the red dashed line is the fitting result of TeV light curve presented in \citep{lhaaso2023tera}. The cyan dashed line is the result of fitting the TeV light curve using the keV-MeV cumulative curve, and the corner plots of this fitting are shown on the right. The residuals for these fittings are shown in panel (b). The keV-MeV flux light curve measured with GECAM-C \citep{HXMT-GECAM:GRB221009A} is displayed in panel (c). } 
  \label{fig:fitting}
\end{figure}

Based on the comprehensive measurements of the prompt keV-MeV and TeV light curve with GECAM-C \citep{HXMT-GECAM:GRB221009A} and with LHAASO \citep{lhaaso2023tera} respectively, we can explore the relationship between the keV-MeV and TeV emission.
As shown in Figure \ref{fig:flux}, we notice that there is an interesting similarity between the cumulative curve of the prompt keV-MeV emission and the TeV light curve. This similarity is also confirmed by the fact that the cumulative curve of the keV-MeV emission can well fit the rising phase of the TeV light curve by adding a time shift, as shown in Figure \ref{fig:fitting}. 

The TeV and keV-MeV light curves are plotted in panel (a) and panel (c) of Figure \ref{fig:fitting}, respectively. There are six time intervals defined for convenience: A for \T+(227,230)\,s, B for \T+(230,235)\,s, C for \T+(235,245.5)\,s, D for \T+(245.5,258)\,s, E for \T+(258,270)\,s and F for \T+(270,300)\,s.

First, we reproduced the fit results of the TeV light curve presented in \cite{lhaaso2023tera}. Since the time of interest for the present work is from \T+230 s to \T+300 s, we just showed two power-law (PL) components which could fit the overall shape of the light curve. The fit result and its residuals are shown in red dashed lines in panel (a) and panel (b), respectively. The index of the PL to fit the rising of TeV light curve is about 1.8. We stress that, when extrapolate this PL component to early time, there is very large and significant deviation (paucity in the observed flux) in the time interval A for this PL component compared to the observation data. We note that, adding an extra PL component to fit the rapid rise at the initial stage of the TeV light curve will not significantly change the PL component with index of 1.8, thus cannot solve this problem. 

Inspired by the similarity mentioned above, we tried to fit the TeV light curve using the cumulative keV-MeV light curve, which can be described by:
\begin{equation}   % fitting Model
f(t,a,b)=10^a \cdot \mathcal{C}(t,b),
\label{equ:CDF_Fitting}
\end{equation}
and
\begin{equation}   % fitting Model
\mathcal{C}(t,b) =\int_{t_{0}}^{t}K(\tau-b)d\tau,
\label{equ:cdf}
\end{equation}
where $t_0$ is the start time of the cumulative calculation, $b$ is the time shift (in units of second), $K$ is the keV-MeV light curve, $\mathcal{C}$ is the cumulative keV-MeV light curve, $a$ is the normalization factor to fit $\mathcal{C}$ to the TeV light curve.
Note that the lower limit of the integration time for $\mathcal{C}$ is the beginning of the first main bump (i.e. \T+225\,s), and the time range of the fit is from 235 to 245.5\,s (i.e. time interval C) without the data in time interval B to avoid the influence of the rapid rise in this time range. 

The \texttt{emcee v3.1.1} \citep{Foreman_Mackey_2013} package was used to do this fitting with the \texttt{MCMC} method.
The posterior distribution of the fitted results are shown in the right panel of Figure \ref{fig:fitting}. It can be seen that all parameters are well constrained. The normalization amplitude constant is $a=-4.00^{+0.01}_{-0.01}$, and the time delay of TeV light curve relative to the cumulative keV-MeV light curve is $b=4.45^{+0.26}_{-0.26}$\,s. 

Interestingly, our cumulative fit (cyan dashed line) could well track the rising curve of TeV emission from the very beginning to the peak, and the problem of the large deviation in time interval A of the PL fit (red dashed line) could be naturally eliminated, as shown in panel (a) of the Figure \ref{fig:fitting}. The residuals in time interval C is also improved compared to the PL fit. Moreover, we find that, for this cumulative fit (cyan dashed line), the structures of the excess residuals in time interval B (panel b) track the light curve of prompt keV-MeV emission (panel c). We also note that there is a modest excess (from about \T+260\,s to 270\,s) of the TeV light curve in time interval E, which again generally follow the keV-MeV light curve of the second bump of main burst. 

Based on the above results, we find that the cumulative keV-MeV light curve could well explain the major component of the TeV light curve in the rising stage before the peak. 
Moreover, both the rapid increase in time interval B and the modest excess in time interval E track the light curve bumps of the prompt keV-MeV emission. The simultaneity implies an interpretation of TeV excesses by the IC scatterings of the prompt emission off the energetic electrons in the external shock (see Section \ref{sec:emission}).

\section{Physical modeling and implication}
We provide physical interpretation to the findings in Section \ref{sec:data}. The correlation of the cumulative keV-MeV and TeV emission and the relative TeV time delay may reveals that the GRB source is continuously powering the TeV emission. Consider a physical picture that the central engine continuously injects energy into the external shock with ejecta, where the keV-MeV activity reflects the central engine energy release, and the time delay is naturally introduced as the ejecta takes time to catch up with the external shock. We can derive the Lorentz factor of ejecta by the time delay (Section \ref{sec:shock}), and model the TeV data well with IC emission, accounting for both the correlation and the TeV excess (section \ref{sec:emission}).

\subsection{Shock dynamics with continuous energy injection}
\label{sec:shock}

Consider that the kinetic luminosity released from the central engine is $L_k(T) = (1/\eta_\gamma-1)L_\gamma(T)$, where $\eta_\gamma$ is the radiative efficiency, assumed to be constant, $T$ is the observer time, with $T=0$ for the launch of the jet. Thus the temporal evolution $L_k(T)$ follows the prompt keV-MeV light curve $L_\gamma(T)$.
The kinetic energy is carried by the ejecta of velocity $v_e$, and will eventually collide into the decelerating shocked fluid of velocity $v$ at radius $R$ (from the central engine at the origin) and time $t$ after a delay time of $R/v_e$ in the {\it unshocked medium frame}. 
Therefore, the rate of energy injection by the ejecta into the shock is $dE/dt = (1 - v/v_e) L_k(t - R/v_e)$. By noting that $dT/(1+z) = (1-v/c)dt$ and $T/(1+z) = t - R/c$, the energy injection rate in the observer's time is $dE/dT \propto L_k(T-\Delta T)$, %$ \equiv L_k[T - (1+z) (1-v_e/c) R/v_e]$. 
with $\Delta T$ being the observed time lag between the ejection of an ejecta element from the central source and the injection of its energy to the shock, corresponding to an observed delay between the prompt and external-shock emission:
\begin{align} \label{eq:DT}
    \frac{\Delta T(T)}{1+z} \equiv \left(1-\frac{v_e}{c}\right) \frac{R(T)}{v_e} \ .
\end{align}
The observed delay resembles its form $R/v_e$ in the unshocked medium frame, but with an additional aberration factor due to the relativistic motion of the ejecta.

\begin{figure}[!t]
  \centerline{
      \includegraphics[width=0.45\columnwidth]{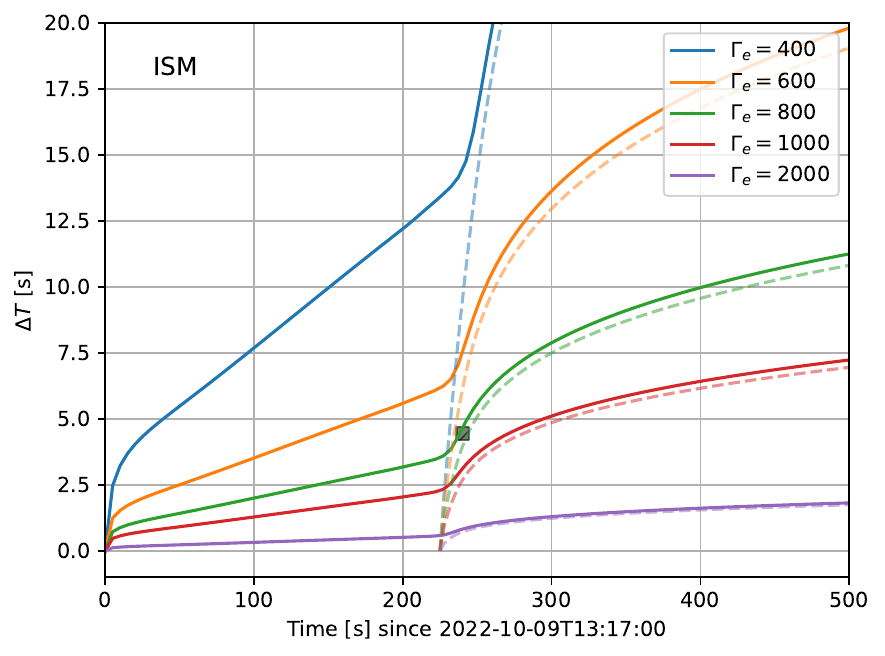}
      \includegraphics[width=0.45\columnwidth]{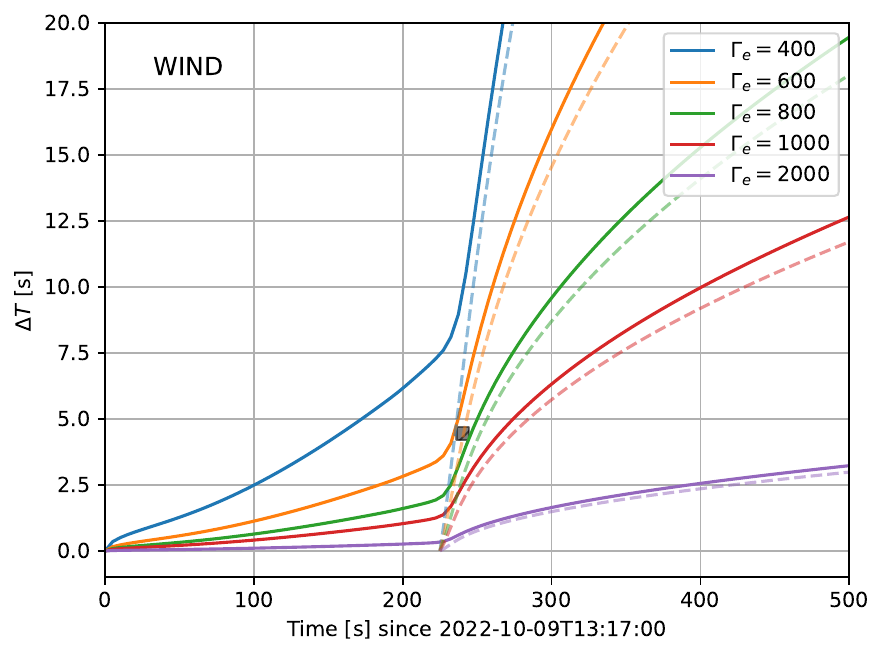}
      }
  \caption{Evolution of delay time induced by kinetic energy propagation for ISM (left) and wind (right) models. The shaded square corresponds to $T_C = T_{\rm ref} + (235, 245.5)\,{\rm s}$ and $\Delta T_C = 4.45^{+0.26}_{-0.26}\,{\rm s}$, the fitted time period and time translation between keV-MeV and TeV emission in Figure \ref{fig:fitting}. Throughout all the models, we have chosen $T_0 = T_{\rm ref}+0$\,s (solid lines) and $T_0 = T_{\rm ref}+225$\,s (dashed lines) and the fiducial parameters as $\eta_\gamma = A_0 = A_{34} = 1$.}
  \label{fig:DT}
\end{figure}

Since $\Delta T \propto R$, the delay is expected to grow over time as the shock radius expands, and its specific evolution is governed by the shock dynamics $v(E, R)$. 
Note that unlike the traditional GRB afterglow models where all energy is instantly released, in our model $E$ is continuously powered by the kinetic ejecta ($dE/dt$) from zero. Therefore, it is crucial to describe $v(E, R)$ near $E=0$, where the energy of the shocked ejecta could be significant.
Consider that the shock energy is the combined kinetic and thermal energy of the shocked ejecta and swept-up medium, $E = (\Gamma\bar\Gamma-1)M_0 c^2 + (\Gamma^2-1)M c^2$, where $M_0c^2 \equiv E/(\Gamma_e-1)$ and $Mc^2 \equiv 4 \pi R^{3-k} A m_p c^2 / (3-k)$ are the rest mass energies of the ejecta and medium, $\bar\Gamma \simeq (\Gamma/\Gamma_e + \Gamma_e/\Gamma)/2$ is the relative Lorentz factor of the shocked fluid (where $\Gamma \equiv 1/\sqrt{1-v^2/c^2}$) measured in the frame of the unshocked ejecta (where $\Gamma_e \equiv 1/\sqrt{1-v_e^2/c^2}$).
The medium number density $n=AR^{-k}$ is given by  parameter $A$ and index $k$ ($k=0$ for ISM and $k=2$ for wind). 

Given the shock dynamics above, we can derive $\Gamma_e$ based on the fitted time delay $\Delta T_C = 4.45^{+0.26}_{-0.26}\,{\rm s}$ for time interval C: $T_C = T_{\rm ref} + (235, 245.5)\,{\rm s}$ (Figure \ref{fig:fitting}). 
First, we give an analytical estimation. Approximating the shock dynamics as $R=\alpha\Gamma^2cT$, where coefficient $\alpha$ represents the uncertainty due to diverse energy injection histories, together with $E\simeq\Gamma^2Mc^2$, the total injected energy to the shock up to time $T$, one derives $R \simeq [\alpha(3-k) T E/(4 \pi A m_p c)]^{1/(4-k)}$. Note $\Delta T/(1+z) \simeq R/(2\Gamma_e^2 c)$, if $\Gamma_e \gg 1$. We give the expression of $\Gamma_e$ in terms of $T$ and $\Delta T$:
\begin{align} \label{eq:g0}
    \Gamma_e \simeq \left[ \frac{\alpha(3-k) T E_\gamma/\eta_\gamma}{2^{4-k} 4 \pi A m_p c^{5-k} \Delta T^{4-k}} \right]^{1/(8-2k)} \simeq 
    \begin{cases}
    740\,(\eta_\gamma A_0/\alpha)^{-1/8} &\quad ({\rm ISM}) \\
    680\,(\eta_\gamma A_{34}/\alpha)^{-1/4} &\quad ({\rm wind})
    \end{cases}
    \ ,
\end{align}
where $E \simeq E_\gamma/\eta_\gamma$, $A_0=A/1\rm cm^{-3}$ and $A_{34}=A/10^{34}$cm$^{-1}$. 
For numerical values, we take $\Delta T=\Delta T_C$ at $T=T_C-T_0\approx15$s, where $T_0 = T_{\rm ref}+225\,$s is adopted for reference since the majority of energy is released thereafter. But the result is very weakly dependent of the parameter uncertainties, including $T_0$.

Next, in the more careful numerical treatment of the shock dynamics, as described above, with the energy injection following exactly the prompt keV-MeV light curve, we derive $\Gamma_e$ by $\Delta T - T$ diagram as shown in Figure \ref{fig:DT}. 
Note that alongside the case of $T_0=T_{\rm ref}+225\,$s, we also show a more physical choice of $T_0=T_{\rm ref}$, where the precursor of GRB 221009A emerged \citep{HXMT-GECAM:GRB221009A}. The results of $\Delta T$ after $T_{\rm ref}+225\,$s become similar between the two choices of $T_0$, indicating the weak dependence of the dynamics on the $T_0$ value.  More importantly, the estimate of Eq. (\ref{eq:g0}) appears to be well consistent with the results of numerical modeling. Thus, thanks to the direct measurement of $\Delta T_C$ (Figure \ref{fig:fitting}), $\Gamma_e$ can be determined with weak dependence on parameter uncertainties (Eq.\ref{eq:g0} and Figure \ref{fig:DT}).

\begin{figure}[!t]
  \centerline{
      \includegraphics[width=0.45\columnwidth]{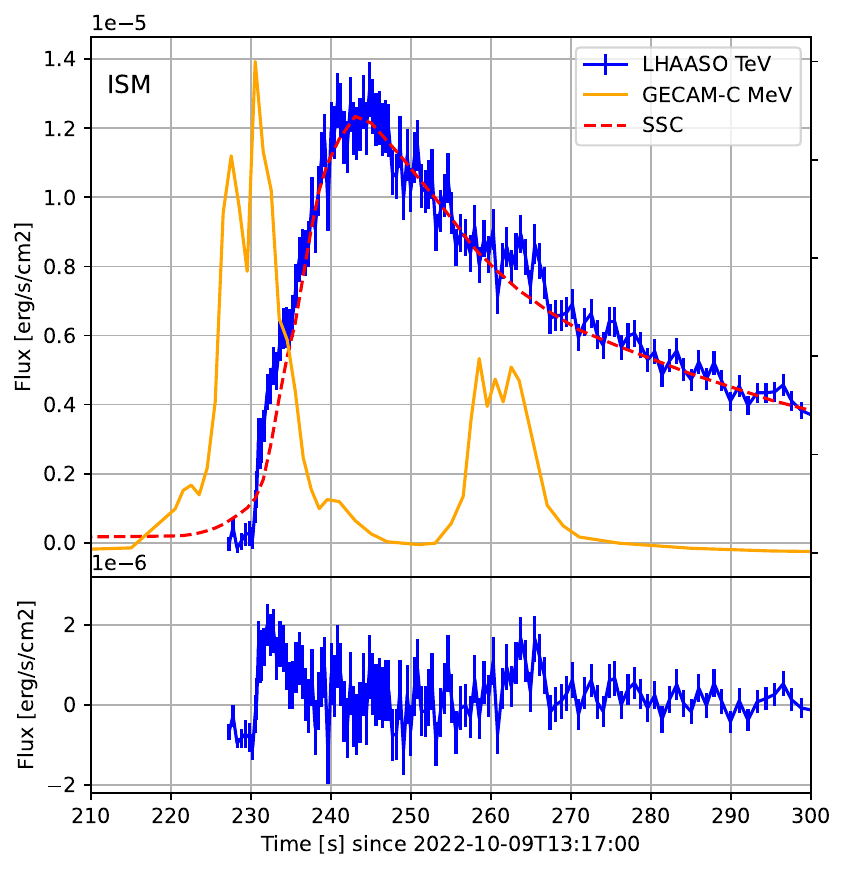}
      \includegraphics[width=0.45\columnwidth]{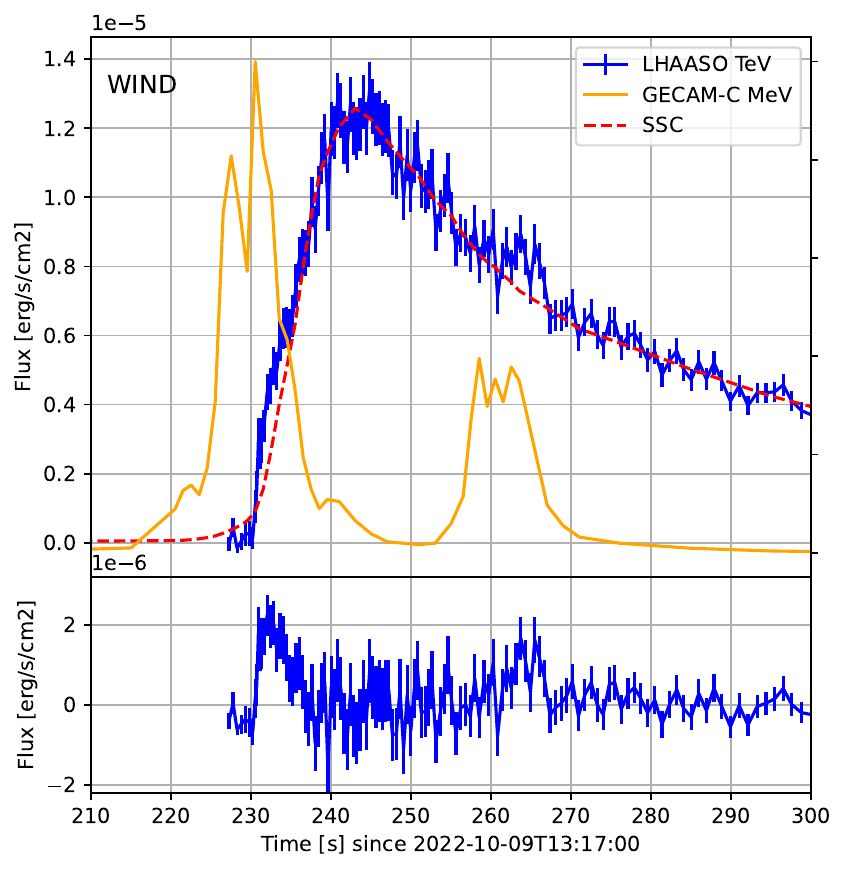}
      }
  \centerline{
      \includegraphics[width=0.45\columnwidth]{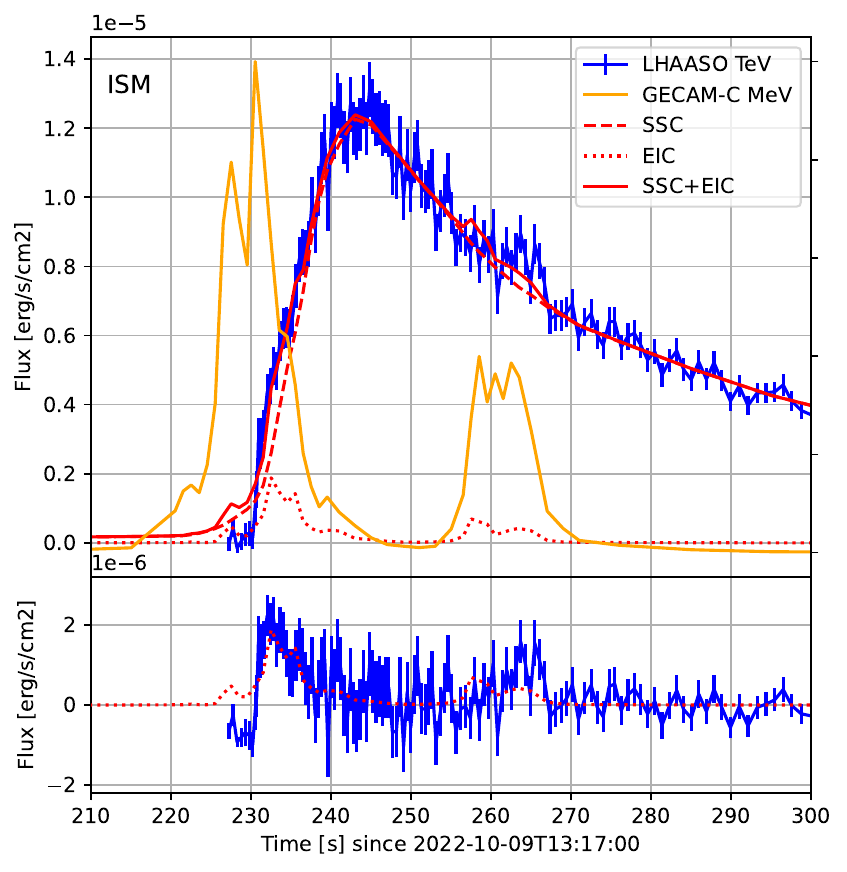}
      \includegraphics[width=0.45\columnwidth]{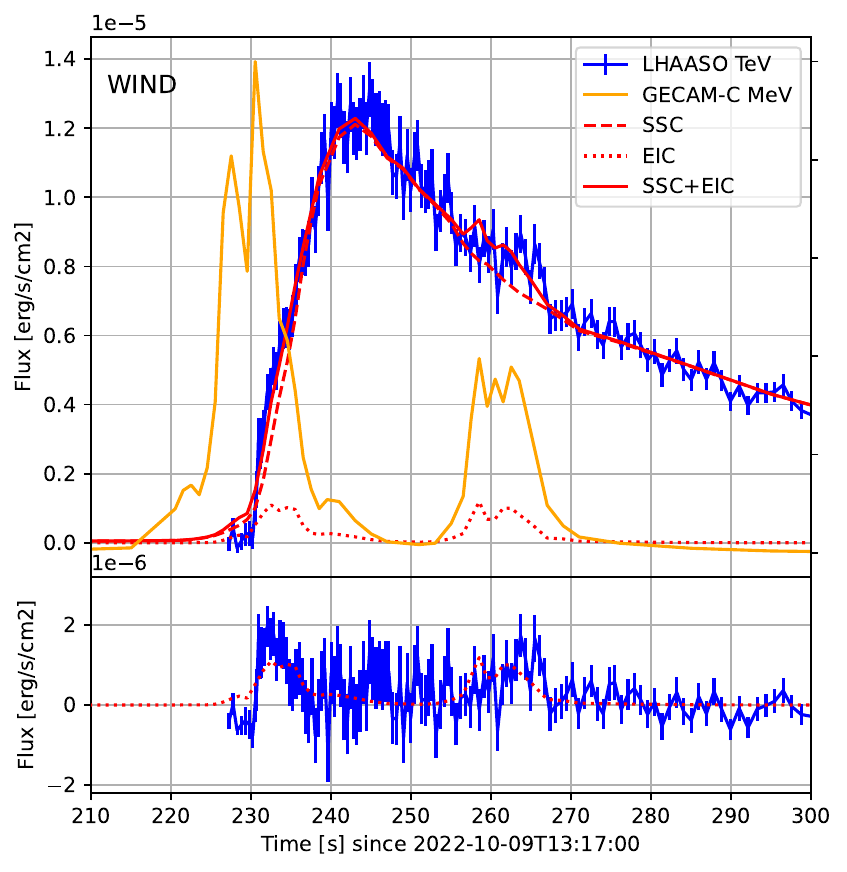}
      }
  \caption{The best-fit SSC/(SSC+EIC)-emission and ISM/wind-medium solutions (red lines) to the LHAASO TeV light curves (blue) in upper panel for each subplot. The data residuals (blue) for the SSC component are shown together with the EIC component (red dotted) in lower panels. 
  The GECAM-C fluxes (orange) with arbitrary normalization are shown in the background for the time reference. 
  The model fittings were performed over \T+(230, 300)\,s.
  }
  \label{fig:modeling}
\end{figure}

\subsection{Synchrotron self-Compton and external inverse-Compton emission}
\label{sec:emission}

In order for a qualitative explanation of the correlation between keV-MeV and TeV emission, we first analytically derive the synchrotron self-Compton (SSC) flux scaling, following \cite{1998ApJ...497L..17S,2001ApJ...548..787S}. The SSC emission by the shocked electrons is dominated by electrons with either the minimum ($\gamma_m$) or cooling Lorentz factor ($\gamma_c$), corresponding to fast or slow cooling regime, respectively. But for either case the SSC luminosity can be given by $L_{\rm SSC}\propto\Gamma^2N_e\gamma_m\gamma_c\eta U_e/(1+x)$, provided that the number-distribution index of injected electrons is $p\ga2$, where $N_e$ is the total number of shocked electrons, $x=L_{\rm SSC}/L_{\rm syn}$ is the luminosity ratio, and $\eta=\min[(\gamma_c/\gamma_m)^{2-p},1]$ is the radiation efficiency of the injected electrons. The value of $\eta$ is almost a constant of $O(1)$ if $p\sim2$, hence $1+x$, as function of $\eta$, is also constant roughly. Note that $E\propto\Gamma^2N_e$, $\gamma_m\propto\Gamma$, $\gamma_c\propto\Gamma/[RU_B(1+x)]$, and the ratio of electron to magnetic-field energy density $U_e/U_B=\epsilon_e/\epsilon_B$ is constant. Putting all together, the SSC flux in the spectral segment with photon index $\ga2$ is $\nu F_\nu^{\rm SSC}\propto L_{\rm SSC}\propto E\Gamma^2/R\propto E/(\alpha T)$. This scaling holds in general.

As for GRB 221009A, a precursor occurs at $T_{\rm ref}$, 230s ahead of the TeV rise, which lasts for only $\sim10$s, thus $T$ does not increase significantly during the TeV rise, i.e,  $\nu F_\nu^{\rm SSC}\propto E$ is roughly available. Moreover, the early start time at the precursor also leads to a smaller change of $\Delta T$ during time interval C (Figure \ref{fig:DT}), aligning more closely with the results of a constant time lag presented in Figure \ref{fig:fitting}. So, the TeV flux should follow the cumulative keV-MeV flux in GRB 221009A. This is largely because of a strong and early precursor. 

In the following, we carry out numerical modelling of data, in order for a careful interpretation of the correlation between keV-MeV and TeV emission. 
We fix the start time $T_0=T_{\rm ref}$ hereafter. Given the shock dynamics with energy injection following $L_\gamma(T)$ of GRB 221009A, starting from the precurosr, we calculate SSC radiation from the external shock, taking into account Klein-Nishina effect and photon-photon annihilation from source photons and extragalactic background lights.
We fit the TeV data with 6 parameters: $\eta_\gamma$, $A$, $\Gamma_e$, $\epsilon_e$, $\epsilon_B$, $p$. The fittings are performed over \T+(230, 300)\,s.
This gives $119$ data points and thus $\nu = 119-6=113$ fitting degrees of freedom.
We do not include jet structure in the calculation, since we find that off-axis geometry does not significantly improve the light curve fitting.

The SSC emission can well fit the TeV rise and the light curve peak (top panel of Figure \ref{fig:modeling}). However, similar to the finding of \cite{2023arXiv231201447D}, we find that the minimalist energy injection model fails to explain the TeV data as it produces a conspicuous bump following the second prompt bump.
Here for simplicity in the calculation, we just set a much lower energy injection efficiency $\epsilon$ for the second prompt peak ($L_k = \epsilon_{1,2}(1/\eta_\gamma-1)L_\gamma$). 
By setting $\epsilon_2/\epsilon_1 \lesssim 0.01$, the effect of the second injection becomes negligible (see Section \ref{sec:discussion} for other discussion).
As a result, SSC model yields a good fit to the TeV data, with reduced chi-squared $\chi_\nu^2 = 1.78$ for the ISM-type medium ($k=0$), and $\chi_\nu^2 = 2.01$ for the wind-medium ($k=2$), as shown in the top panel of Figure \ref{fig:modeling}.

As discussed above, we find regions where there are excesses in the TeV light curve with respect to the SSC model that are well aligned with the keV-MeV light curve. 
We note that \cite{2023arXiv231201447D} interpreted the excesses as reverse shock emission. 
However, we indicate that any kind of emission powered by kinetic energy injection should suffer from time delay that increases with time by Eq.(\ref{eq:DT}) (see also Figure \ref{fig:DT}), and hence the excesses caused by injection cannot be well aligned in time with the prompt activities. One exception of $\Delta T = 0$ is when $v_e = c$, i.e. the emission is directly powered by the prompt photons, and a straightforward interpretation would be the IC scattering of the prompt photons.

Here we calculate the external inverse-Compton (EIC) emission using the prompt photons as the seed photons to be scattered. At the site of the external shock, the Lorentz-invariant spectral number density [1/eV/cm$^3$] of the external photons can be estimated by $n_{\rm ext} \simeq L_{\rm ext}/(4\pi R^2 c)$, where $L_{\rm ext}$ is the count-rate luminosity of external emission [1/eV/s], and is related to the observed prompt emission $L_\gamma$. 
Unfortunately, the detection energy range (from about 15\,keV to 6\,MeV) of GECAM-C is in the Klein-Nishina regime, and therefore we lack the direct measurement of the low-energy spectrum of the prompt emission that contributes most to the EIC flux.

For a simple estimation, we assume that the low-energy prompt spectrum that contribute the most EIC flux is proportional to that of the spectrum measured with GECAM-C. This will introduce one additional degree of freedom in the model fitting. 
Nevertheless, we find that the SSC+EIC model produce statistically better fit (smaller reduced chi-square) to the excesses compared with the pure SSC model, yielding $\chi_\nu^2 = 1.13$ (ISM) and $\chi_\nu^2 = 1.31$ (wind), shown in the bottom panel of Figure \ref{fig:modeling}.
Note that there is a slight, inevitable excess of model to data just before $T=230\,$s corresponding to the injection of ejecta around $T-\Delta T \sim 225\,$s, where the prompt emission fluxes has risen to a non-negligible level. This discrepancy could potentially be resolved by assuming a temporal evolution of $\Gamma_e$. 

To conclude, the successful modelling of TeV data with energy injection following the keV-MeV light curve further supports that the close relation between keV-MeV and TeV emission reveals the continuous powering of the external shock by the time-varying jet. 

\section{Discussion}\label{sec:discussion}

In data analysis, we find that the cumulative of keV-MeV prompt emission could well fit the rising stage of the TeV light curve, revealing a remarkably close relation between the prompt emission and external-shock emission. Our fit can also naturally eliminate the problem of the paucity of the observed TeV flux in early time which is inevitable in the empirical fitting with power law function (Figure \ref{fig:fitting}). 

Importantly, our results provide a direct evidence of the continuous energy injection to the external shock, as demonstrated by the qualitative discussion and the numerical modelling of the TeV data with energy injection following keV-MeV light curve (Section \ref{sec:emission}). We note that in general it is only expected $L_{\rm SSC}\propto E/T$. The reason that in GRB 221009A the TeV flux follows closely the cumulative prompt emission is the strong and very early precursor, compared to the duration of the main bump of the prompt emission.
Our results also indicate that the efficiency of both the keV-MeV emission and TeV mission relative to the total energy of the jet seems to be nearly constant for this burst, otherwise we cannot observe such a good agreement between the cumulative keV-MeV flux light curve and the rising part of the TeV flux light curve.

It is worth of comparing the continuous injection with the impulsive injection model, usually adopted in literature \citep[see, e.g.,][]{gao2013NewAR}. 
In the impulsive injection model, a start time for the sudden release of a homogeneous ejecta should be set, e.g., either the detector trigger time or some point in the bursting phase \citep[e.g.,][for GRB 221009A]{lhaaso2023tera}. The early afterglow behavior in the model depends on the choice of the start time \citep[e.g.,][]{2007ApJ...655..973K}. However, in the continuous injection case, such a precisely chosen start time is much less important. One only needs to start the integration of energy earlier than the main energy injection. As Figure \ref{fig:DT} shows, different choices of $T_0$ result in similar dynamics in late time. 

The temporal increasing of the SSC flux is determined by the energy injection history, in difference with the impulsive injection model; but the SSC behavior after the time when the energy injection finished, i.e., after the TeV light curve peak, is similar between the two models. Thus, it is crucial to consider continuous energy injection when discussing the early external shock emission, even more seriously in the onset and rising phase of the external shock emission. If the true physics is a continuous injection behind, the impulsive approximation may work for data after the light curve peak by carefully choosing the start time of the shock, but it cannot easily work for the rising phase because it does not catch the physical process. For example, in contrast to Eq. (\ref{eq:g0}), the estimate of $\Gamma_e$ for GRB 221009A in the impulsive-injection model is $\Gamma_e \simeq \left[ (3-k) E_\gamma/[2^{4-k}4\pi A m_p c^{5-k} (T_p-T_0)^{3-k} \eta_\gamma] \right]^{1/(8-2k)} \simeq 435\,(\eta_\gamma A_0)^{-1/8}$ (ISM) and $431\,(\eta_\gamma A_{34})^{-1/4}$ (wind), respectively, where the TeV peak time relative to $T_0$, $T_p-T_0\approx15$s, is used. The result is different from that of Eq. (\ref{eq:g0}), and depends on the adopted $T_0$ value. 

Although the majority of the TeV flux could be fit with cumulative keV-MeV light curve, there are excesses in the TeV light curve (Figure \ref{fig:fitting}). One is the rapid increase at the beginning, and the other is the modest significant excess from \T+260\,s to 270\,s. We find that both excess generally and contemporarily tracks the pulses in the keV-MeV light curve, indicating that these excess components are probably from the IC scattering of the prompt photons by the energetic electrons in the external shock. This has been demonstrated by our data modelling with EIC contribution (Fig \ref{fig:modeling}).

According to our calculation, the soft X-ray photons in the prompt emission is the main part of the EIC seed photons. However, the direct observation in soft X-ray band is missing for the prompt emission of this burst. 
We assume that the soft X-ray light curve resembles that of the hard X-ray measured by GECAM-C \citep{HXMT-GECAM:GRB221009A}.
Note that the rapid increase in the initial stage of the TeV light curve is hard to be interpreted in other models \citep[e.g., the impulsive injection model in][]{lhaaso2023tera}, but it could be reasonably explained by the EIC emission in our model, because there is also a rapid rising in the prompt keV-MeV light curve at the same time.

Finally, we note that the second bump (from \T + 250\,s to 270\,s) of the main burst in the keV-MeV light curve apparently has much less influence to the shock dynamics and hence the TeV emission compared to the first bump (from \T + 225\,s to 240\,s). This may imply that the physical properties corresponding to the second bump is different from that of the first bump. For example, the ejecta Lorentz factor $\Gamma_e$ for the second bump could be much smaller, thus the energy injection to the shock occurs much delayed, $\Delta T\approx R/2\Gamma_e^2v_e$ (eq. \ref{eq:DT}), and the duration of energy injection would be prolonged. Additionally, the relatively larger radius leads to larger spreading time ($T_{\rm spr}\sim R/2\Gamma^2c$) which further smooths out the light curve. So there could be no obvious bump in the TeV light curve.

\section{Conclusion}

By analyzing the unprecedented observation data of the brightest GRB 221009A made with GECAM-C \citep{HXMT-GECAM:GRB221009A} and LHAASO, we find that the keV-MeV and TeV emission are closely related with each other and important physics of this GRB and the relativistic jet are derived by the physical modelling:

(1) The cumulative light curve of keV-MeV emission could well fit the rising stage of the TeV light curve of GRB 221009A, with a time delay of TeV emission. This relation can be interpreted by a continuous-, rather than impulsive-, energy injection model, where the external shock that accounts for the TeV emission is powered by the energy injection following the keV-MeV emission, including the precursor. The rising phase of the external shock emission is especially dependent on the energy injection history.

(2) The measured time delay $4.45^{+0.26}_{-0.26}$\,s of the TeV emission with respect to the cumulative keV-MeV emission provides a direct probe to the jet Lorentz factor of $\sim 740\,(\eta_\gamma A_0/\alpha)^{-1/8}$ (ISM) and $\sim 680\,(\eta_\gamma A_{34}/\alpha)^{-1/4}$ (wind), respectively.

(3) Both the rapid increase in the initial stage and the excess from about \T+260\,s to 270\,s in the TeV light curve are tracking the light-curve bumps in the prompt keV-MeV emission. These TeV excesses can be naturally interpreted by EIC scatterings of the inner-coming prompt emission by the energetic electrons in external shock.

\section*{Acknowledgments}
 We appreciate the reviewer for helpful comments and suggestions which improved this work.
 We thank the support from the National Key R\&D Program of China (Grant No. 2021YFA0718500), % xiong shaolin for HXMT & GECAM
the National Natural Science Foundation of China 
(Grant No. 12273042, % xiong shaolin 
U1931201,% li zhuo
12333007 and 12027803, % Zhang Shuang-Nan) 
and the Strategic Priority Research Program of the Chinese Academy of Sciences (Grant No. XDB0550300, % xiong shaolin for xiandaoB
XDA30050000). % xiong shaolin for GTM
The GECAM (Huairou-1) mission is supported by the Strategic Priority Research Program on Space Science of the Chinese Academy of Sciences (XDA15360000). S.-X.Y. acknowledges support from the Chinese Academy of Sciences (grant Nos. E329A3M1 and E3545KU2). We thank Hao Zhou, Zhiguo Yao, Huicai Li, Zhen Cao for valuable discussion. We appreciate the GECAM and LHAASO teams for their help.

\bibliography{msNote}{}
\bibliographystyle{aasjournal}

\end{document}